\documentclass[aps,prl,twocolumn,groupedaddress,showpacs]{revtex4}
\usepackage{latexsym,amsfonts,amssymb,amsbsy}
\usepackage{graphicx}

\begin{document}

\title{Domain free high field superlattice transport}

\author{T. Feil}
\affiliation{Institut f\"ur Experimentalphysik, Universit\"at Regensburg, 93040 Regensburg, Germany}

\author{H.-P. Tranitz}
\affiliation{Institut f\"ur Experimentalphysik, Universit\"at Regensburg, 93040 Regensburg, Germany}

\author{M. Reinwald}
\affiliation{Institut f\"ur Experimentalphysik, Universit\"at Regensburg, 93040 Regensburg, Germany}

\author{W. Wegscheider}
\affiliation{Institut f\"ur Experimentalphysik, Universit\"at Regensburg, 93040 Regensburg, Germany}

\date{\today}

\begin{abstract}
The fundamental condition for operating a superlattice (SL) as a Bloch oscillator is a homogeneous electric field alignment under large applied bias. This can be done by combining the nonlinear miniband transport channel with a parallel shunt channel, connected through lateral transport. We report here about the first realization of such a SL-shunt system. With the cleaved-edge-overgrowth (CEO) method we combine an undoped superlattice, which acts as the shunt, with a two-dimensional (2D) cleaved-edge surface superlattice channel. Our results confirm the long predicted current-voltage-characteristic (I-V) of a SL in a homogeneous DC electric field. 
Excitation experiments in the GHz regime show that our shunt approach also works in the presence of an external AC electric field. The device concept opens a clear path towards the realization of a solid state Bloch oscillator. 
\end{abstract}


\maketitle

\section{}
\begin{figure}
  \includegraphics[width=0.95\columnwidth]{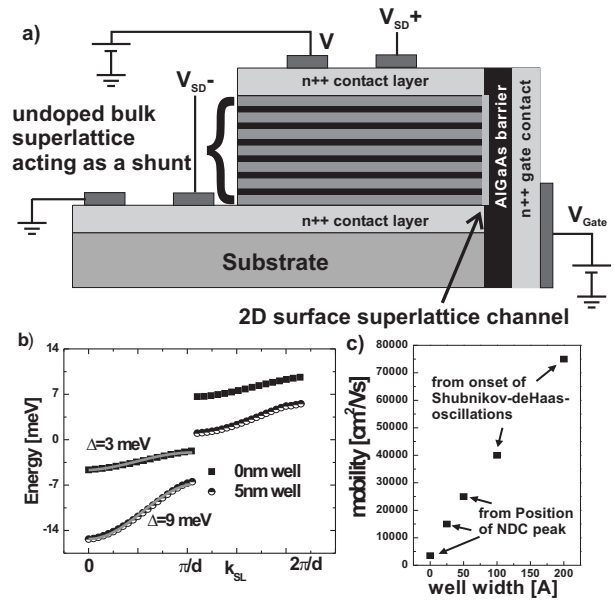}
  \vspace{-0.5cm}
  \caption{(a) Schematic drawing of the sample structure. In (b) the theoretically calculated bandstructures along the surface SL are plotted for 0 nm and 5 nm cleaved-edge well widths, together with a simple cosine fit. (c) An increase in the cleaved-edge well widths leads to a strong increase in the mobility of the surface SL.}
  \label{sample}
  \vspace{-0.2cm}
\end{figure}

\begin{figure}
  \vspace{-0.7cm}
  \includegraphics[width=1.0\columnwidth]{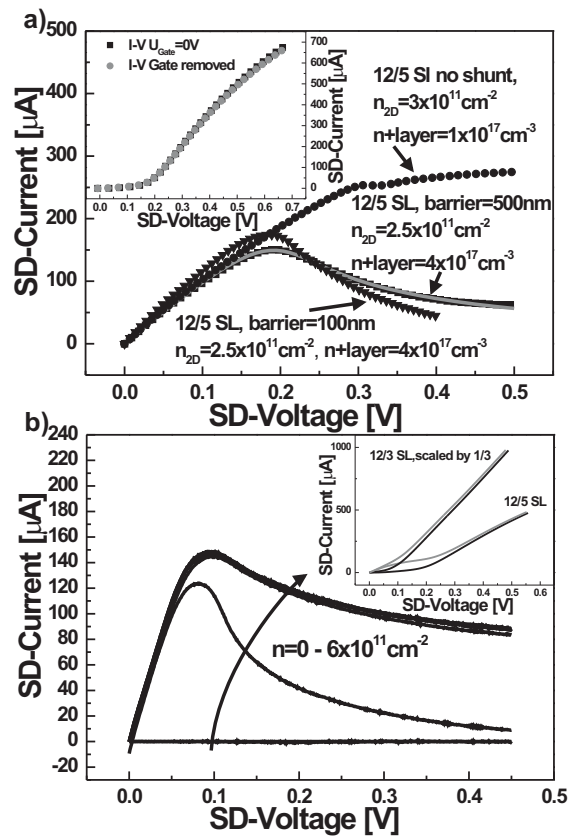}
  \vspace{-0.8cm}
  \caption{(a) Current through the surface SL of different 12/5 SL structures. The dots show a shunt free system which results in an inhomogeneous electric field distribution. The squares show the transport in a structure with a 500nm wide gate barrier. The resulting I-V-characteristic agrees excellently with the semiclassical fit (grey line). A thinner gate barrier (triangles) leads to quenching effects for increasing SD-voltage. (b) Current through the surface SL of a 12/3 SL sample for channel densities from $0$ to $6 \times 10^{11} cm^{-2}$ in steps of $10^{11} cm^{-2}$. The observed saturation in transport is theoretically expected.}
  \label{IVs}
  \vspace{-0.2cm}
\end{figure}

\vspace{-1cm}
When Esaki and Tsu \cite{esaki70} proposed the fabrication of SL structures, they showed that these should exhibit a large region of negative-differential conductance (NDC) due to the occurance of Bloch oscillations \cite{bloch28}, \cite{zener34}. Since they found Bloch frequencies in the terahertz (THz) range, they suggested to use the SL as a solid state source in this regime. Semiclassical calculations of the dynamic conductivity of a SL by Ktitorov {\it et al.} \cite{KtSiSi1972} confirmed that the SL acts as a gain medium for all frequencies up to the Bloch frequency. About twenty years after these initial predictions, experiments, successfully revealing the formation of Wannier-Stark ladders in biased SLs \cite{mendez88}, and resolving Bloch oscillations with the help of time domain spectroscopy \cite{FeLeShMiCuScMePlScTh1992Oio}, \cite{waschke93}, \cite{dekorsy94}, were performed. The excitation of a biased doped SL \cite{unterrainer96} also showed the  theoretically predicted THz resonances in the photoconductivity of the SL sample.
\begin{figure}
  \vspace{-1cm}
  \includegraphics[width=0.90\columnwidth]{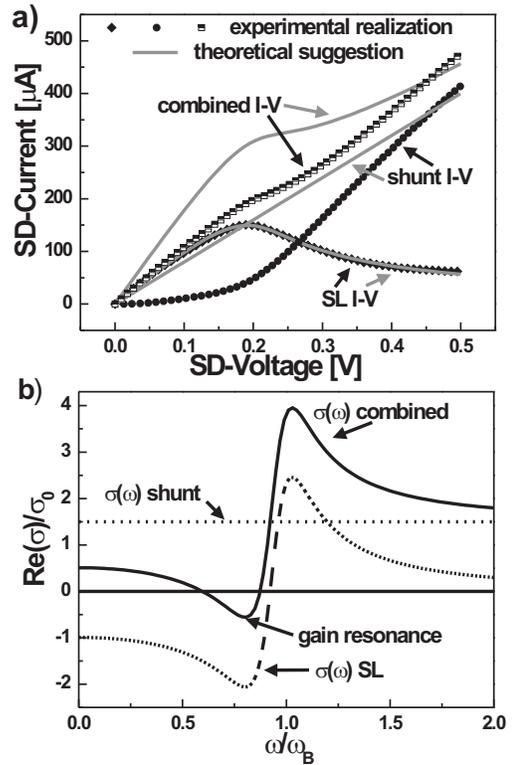}
  \vspace{-0.7cm}
  \caption{(a) Current-Voltage-characteristics of the shunt, the SL and the combined system suggested in \cite{DaGiScAl2003SoE} compared with the experimental realization. Adding the shunt to the superlattice makes the overall conductivity positive and no instabilities occur. (b) If the real part of the dynamic conductivity is smaller than the maximum gain at the resonance of the SL, then the combined system still exhibits gain close to the Bloch frequency; adapted for our samples from \cite{DaGiScAl2003SoE}.}
  \label{Danielproposal}
  \vspace{-0.5cm}
\end{figure} 
Theoretical findings by Willenberg {\it et al.} \cite{WiDoFa2003Igi} extend the Ktitorov results to a wide range of SL parameters and electric field and temperature regions. It was also possible to confirm the theoretically predicted shape of the gain curve, by deriving the dynamic conductivity from time domain spectoscopy results \cite{ShHiOdCh2003}. Recently, a direct observation of the crossover from loss to gain in a biased super-SL sturcture was successfully performed \cite{SaKoLeAl2004}, measuring the change in transmission of a super-SL array, when bias was applied to the structure during the THz excitation.\\
Nevertheless there exists no successful realization of an active solid state Bloch oscillator up to now. The main obstacle in achieving such a device, is the instability of the electric field in a region of NDC. This phenomenon is well known from the Gunn effect \cite{KR1972GeB}. Any fluctuation in the electron density along the device, when it is biased in the NDC region, leads to a growth of the fluctuation and a redistribution of the electric field in domains of small and large magnitudes. This destroys the equidistant Stark-ladder. In contrast, the homogeneous electric field along the device is an essential requirement present in all theoretical calculations predicting gain in a SL structure. Consequently, in order to realize an active Bloch oscillator device one has to overcome the problem of domain formation. One such attempt is the super-SL structure employed in \cite{SaKoLeAl2004}, in which individual SL sections are so short that full grown domains cannot be formed. Under THz excitation this resulted in a homogeneous field distribution across about 75 percent of the device. Another approach towards avoiding the problem of instabilities was theoretically suggested by Daniel {\it et al.} \cite{DaGiScAl2003SoE}. The authors propose to combine the SL with a side shunt resistor which acts as an electric field stabilizer in the static transport regime. Their proposal is summarized in figure \ref{Danielproposal}. The addition of the shunt can make the overall conductivity positive in the regime of NDC. Numerical simulations show that the shunt can stabilize the electric field distribution inside the SL up to a length scale of a few microns. In the dynamic transport regime one tries to arrange the structure in a way, that the real part of the dynamic conductivity of the shunt increases the combined conductivity, but is small enough such that the gain resonance of the SL becomes smaller but is still present, as drawn in figure \ref{Danielproposal}b.\\
In this letter we report the first successful realization of a SL-shunt system that provides a homogeneous field distribution across the SL. Molecular beam epitaxial growth on the sidewall of an undoped SL employing the CEO method \cite{stormer91} allows us to combine the shunt and SL with atomic precision and without problems arising from poorly defined or contaminated interfaces. Since our geometry also results in direct control of the SL channel density, a complete separation of the two transport contributions is possible. Thus, for the first time we can experimentally confirm the long and often predicted I-V-characteristic of a superlattice under stable high field conditions. The device gives entrance to a theoretically well studied, but experimentally so far inaccessible transport regime. The response of the SL-shunt systems to an external high-frequency (HF) field confirms the formation of a domain free field distribution.\\
We realize the SL-shunt system with the CEO method which, in the GaAs/AlGaAs system, allows the atomically precise combination of two different growth steps (figure \ref{sample}(a)).
In a first molecular-beam-epitaxy process we grow an undoped GaAs/Al$_{0.3}$Ga$_{0.7}$As SL structure in between two $n+$-GaAs layers on a semi-insulating [001] GaAs substrate. The undoped SL, whose transport properties can be modified by the doping concentration in the $n+$-layers, acts as the shunt in our structure. This layer sequence is then cleaved in situ along the [110] direction and is overgrown with a GaAs well, followed by an Al$_{0.3}$Ga$_{0.7}$As barrier and a $n+$ GaAs layer acting as a gate contact for the surface superlattice channel. The thickness of the cleaved edge well determines the bandstructure properties and mobilities, whereas the thickness of the gate barrier controls the ratio between source-drain-voltage (SDV) and gate bias. Two different SL systems with 25 periods of 15 nm length and 60 periods of 17 nm length were used for the samples. In both cases the SL well width was 12 nm, while the barriers were 3 nm and 5 nm. For further reference we will call these 12/3 SL and 12/5 SL structures. 
The cleaved edge well thickness was varied between 0 nm and 20 nm. Those, with 5 nm well width and a gate barrier thicknesses of 100 nm and 500 nm, were chosen for the high-field transport studies. Magnetotransport measurements revealed an adjustable 2D channel density between $0$ and $6 \times 10^{11} cm^{-2}$ for gate voltages between $0$ and $7$ V (500 nm barrier width). Figure \ref{sample}(b) shows the bandstructure of the 12/3 SL structure for cleaved edge well widths of 0 nm and 5 nm, calculated with the self-consistent Schr\"odinger-Poisson-Solver nextnano3 \cite{nextnano}. Our calculations show an increasing minibandwidth for thicker wells, but at the same time the inflection point moves towards the brillouin zone boundary and almost completely disappears for the 20 nm well. These findings are in agreement with differently based numerical calculations for equivalent structures \cite{GrVaBaMiHa2003Qrt}. Although the gaps to higher lying states appear rather small, the simulations show that these states belong to minibands formed from excited subbands of the gate induced confinement potential. Since the spatial overlap between these subbands is vanishingly small, no signigicant tunneling between the lowest lying miniband and the excited ones is expected and also not observed in the experiment. The miniband width of the 12/5 SL with the 5nm cleaved-edge well is 3.5 meV\\
The I-V-characteristics of our devices depicted in figure \ref{IVs} are measured at liquid helium temperature. An increasing voltage $V$ is applied between the contact layers of the SL, while the current through the device and the voltage drop $V_{SD}$ across the sample are measured as indicated in figure \ref{sample}(a). In the inset of figure \ref{IVs}(a) a comparison between the I-V traces taken for zero applied gate bias and after the gate was mechanically removed shows, that the gate control indeed allows us to separate between the two transport channels. 
Since the shunt channel alone already represents transport through an undoped SL, the question of possible instabilities in this channel has to be addressed. The experimental results show no abrupt instabilities up to Stark splittings, the energy associated with a voltage drop per SL period, of more than 25 meV. The inset of figure \ref{IVs}(b) shows the combined results for transport through both the shunt and the surface superlattice. The CEO method leads to a very flexible sample design and allows the realization of different ratios between the magnitude of the two transport channels. While the shunt dominates the shape of the 12/3 SL sample, the contributions are more equal for the 12/5 SL. After subtraction of the shunt transport from the I-V-characteristic of the combined system, the current carried by the surface superlattice is recovered and plotted in figure \ref{IVs}. The trace follows the long and often predicted shape. For small electric fields, transport is essentially ohmic. With increasing bias, the electron distribution starts to feel the non-linearity of the dispersion and the current saturates. Beyond this point electrons increasingly start to Bloch oscillate and the transport gradually decreases. In order to emphasize the importance of the shunt channel, figure \ref{IVs}(a) also shows an I-V-characteristic of a sample with no such stabilizer. This was achieved by reduced doping in the n+ layers, for which a finite minimum voltage is needed to inject carriers into the SL. For this sample a typically observed current plateau due to field inhomogeneties is measured. The instability occurs at a source-drain- to gate voltage ratio of about 5 percent. Such high gate voltages are necessary to avoid a strong influence of the sample bias on the gate induced confinement \cite{FeDeWeRoScBiAbRiKe2004Tiw}. This fact is also stressed by the comparison of the 12/5 SL samples with 100 nm and 500 nm thick gate barriers also shown in figure \ref{IVs}(a). There, the current of the thin gate barrier sample drops quickly towards zero, which is largely caused by the fact that the increasing source-drain voltage reduces the confinement potential at the top contact and therefore causes an increasingly inhomogeneous density and field distribution along the surface SL. The 12/5 SL sample I-V is fitted with the Esaki Tsu prediction $I \sim \frac{\omega_B \tau}{1 + \omega_B^2 \tau^2}$, where the peak current was adjusted to the experimental value and a series resistance of about 500 $\Omega$ was included, which was independently confirmed by magneto-resistance measurements. An almost perfect agreement between experiment and theory is found for this trace, but the simple theoretical result does not work equally well for other densities or the 12/3 SL sample. This is attributed to the fact, that a possible increase in the electronic temperature for large electric fields has to be considered \cite{MiPePaSaMoPl1994Mca} and that the device and measurement setup does not allow a direct determination of the voltage applied to the surface SL. In figure \ref{IVs}(b) we plot the I-V-characteristics of the 12/3 SL sample for densities up to $6 \times 10^{11} cm^{-2}$ in steps of $10^{11} cm^{-2}$. For higher densities a saturation of the overall transport is observed. This is theoretically expected from the filling of the 2D miniband  \cite{FeDeWeRoScBiAbRiKe2004Tiw} and confirms our theoretically calculated bandstructure. Small densities still show a more pronounced current drop due to the density reduction caused by the source-drain voltage increase. \\
From the current peak position scattering times of 350 fs and 500 fs, respectively, are extracted for the 12/5 SL and 12/3 SL structures, assuming $\omega_B\tau=1$ at the peak. Since the miniband width is smaller than the longitudinal optical phonon energy of 36 meV in GaAs, prominently accoustic phonon scattering contributes to the inelastic scattering processes and is responsible for the overall transport along the surface SL. Moreover, as depicted in figure \ref{sample}(c) an increase in the cleaved-edge well width shows a strong increase of the mobility. Since the electronic system is thereby shifted away from the SL structure, this indicates strong elastic scattering by interface roughness, which is also observed in conventional SL transport. This elastic interface roughness scattering is responsible for the relatively short relaxation times measured and, since it strongly dominates over the inelastic scattering, also for an observed large reduction of the peak current \cite{MiPePaSaMoPl1994Mca}.\\
Finally we studied the response of our SL/shunt system to an externally applied HF field $F_{\omega}$. The 12/5 SL sample was mounted inside a waveguide which was fitted with an ajustable short at its end. It was then excited with a Gunn-oscillator operating at 65 GHz, mounted with an adjustable attenuator. During excitation the DC I-V-characteristics were measured.
\begin{figure}
  \vspace{-0.9cm}
  \hspace{-0.6cm}
  \includegraphics[width=1.1\columnwidth]{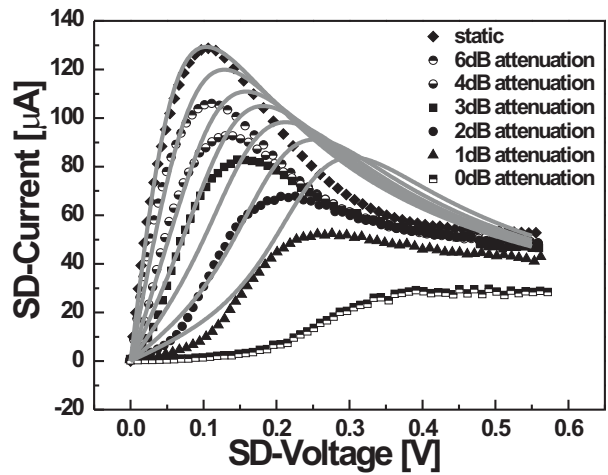}
  \vspace{-0.8cm}
  \caption{The plot shows the DC transport characteristics of a 12/5 SL structure excited at 65 GHz for increasing input power. The grey lines are plotted from the evaluation of a semiclassical theory. The smaller experimental values are attributed to an increased electron temperature not considered in the theory.}
  \label{Gunn}
  \vspace{-0.5cm}
\end{figure}
The results for various excitation powers $P$ are plotted in figure \ref{Gunn}. The device shows a strong response towards the HF excitation. The initially observed NDC peak is completely suppressed for large powers. A comparison with the semiclassically expected response \cite{WiScGrReIgSeRepaKoMeUsIvScKo1997Qad}, which is plotted in 2 dB steps assuming $P \sim F_{\omega}^2$, shows qualitatively the same power dependence as our data. The fact that the measurements give systematically smaller current values than those predicted by theory, is attributed to an increase in the effective electron temperature, as a result of the expected long inelastic scattering times.\\
In summary we present the realization of a SL/shunt system, which allowed us to study static and dynamic transport in surface SLs with a homogeneous electric field distribution, completely avoiding domain formation. The samples reveal the long predicted stable I-V-characteristic of a SL. The large region of NDC due to electrons performing Bloch oscillations is strikingly confirmed. Exciting the samples with GHz radiation also follows the semiclassically expected behavior. Although the response is still quasistatic, due to the short relaxation times, the presented structure is an important first step towards the realization of an acitve solid-state Bloch oscillator. The remaining influence of the gate structure on the measured traces can be further reduced by thicker gate barriers or completely avoided by switching to a modulation doped structure. Moreover, the better achievable mobilities in the 2D surface SL compared to 3D SL transport offer the advantage of a more pronounced gain maximum below the Bloch frequency. 
\begin{acknowledgments}
We would like to thank S. J. Allen for very fruitful discussions and gratefully acknowledge financial support by the DFG via SFB348 and by the BMBF, contract 01BM918. 
\end{acknowledgments}

\end{document}